\documentstyle[twoside,fleqn,espcrc2,epsfig]{article}
\newcommand{\sss}{\scriptscriptstyle}
\title{NLO QCD corrections to hadronic Higgs production with heavy
  quarks} \author{S.~Dawson\address{Physics Department, Brookhaven
    National Laboratory, Upton, NY 11973-5000, USA},
  C.~B.~Jackson\address{Physics Department, Florida State University,
    Tallahassee, FL 32306-4350, USA}, L.~H.~Orr\address{Department of
    Physics and Astronomy, University of Rochester, Rochester, NY
    14627, USA}, \underline{L.~Reina}$^{\rm b}$\thanks{Talk presented by
    L.~Reina.}, D.~Wackeroth\address{Department of Physics, SUNY at
    Buffalo, Buffalo, NY 14260-1500, USA}}
\begin{document}
\begin{abstract}
  The production of a Higgs boson in association with a pair of
  $t\bar{t}$ or $b\bar{b}$ quarks plays a very important role at both
  the Tevatron and the Large Hadron Collider. The theoretical
  prediction of the corresponding cross sections has been improved by
  including the complete next-to-leading order QCD corrections.
  After a brief description of the most relevant technical aspects of
  the calculation, we review the results obtained for both the
  Tevatron and the Large Hadron Collider.
\end{abstract}
\maketitle
\section{INTRODUCTION}
\label{sec:intro}
The existence of a relatively light Higgs boson is both suggested by
precision fits of the Standard Model (SM) and theoretically required
by the Minimal Supersymmetric extension of the Standard Model (MSSM).
Searches at both the Tevatron and the Large Hadron Collider (LHC) will
play a crucial role in testing this hypothesis and in discriminating
between different models that imply the existence of one or more Higgs
bosons.  In this context, the production of a Higgs boson in
association with a heavy quark and antiquark pair, both $t\bar{t}$
and $b\bar{b}$, plays a very important role.

The associated production of a Higgs boson with a pair of $t\bar{t}$
quarks has a very distinctive signature, and can give the only handle on
a direct measurement of the top quark Yukawa coupling, perhaps the
most crucial coupling in exploring the origin of fermion masses.
Observing $p\bar{p}\to t\bar{t}h$ at the Tevatron
($\sqrt{s}\!=\!2$~TeV) will require very high
luminosity~\cite{Goldstein:2000bp} and will probably be beyond the
machine capabilities. On the other hand, if $M_{h}\!\le\!130$~GeV,
$pp\to t\bar{t}h$ is an important discovery channel for a SM-like
Higgs boson at the LHC
($\sqrt{s}\!=\!14$~TeV)~\cite{atlas:1999,Beneke:2000hk,Drollinger:2001ym}.
Given the statistics expected at the LHC, $pp\to t\bar{t}h$, with
$h\to b\bar{b},\tau^+\tau^-,W^+W^-,\gamma\gamma$ will also be
instrumental to the determination of the couplings of a discovered
Higgs
boson~\cite{Zeppenfeld:2000td,Zeppenfeld:2002ng,Belyaev:2002ua,Maltoni:2002jr,Duehrssen:2003at}.
Several analyses show that precisions of the order of 10-15\% on the
measurement of the top quark Yukawa coupling can be obtained with
integrated luminosities of 100~fb$^{-1}$ per detector.  Moreover, the
combined measurements of $pp\to t\bar{t}h$ with $h\to b\bar{b}$ and
$h\to\tau^+\tau^-$ could provide the only model independent
determination of the ratio of the bottom quark to the $\tau$ lepton
Yukawa couplings~\cite{Belyaev:2002ua}.

The associated production of a Higgs boson with a pair of $b\bar{b}$
quarks has a very small cross section in the SM, and can therefore be
used to test the hypothesis of enhanced bottom quark Yukawa couplings
which is common to many extensions of the SM, such as the MSSM for
large values of $\tan\beta$. Both the Tevatron and the LHC will be
able to search for evidence of an enhanced $b\bar{b}h$ production,
looking for a final state containing no bottom quarks (inclusive
production), one bottom quark (semi-inclusive production) or two
bottom quarks (exclusive production). An anomalously large inclusive
Higgs boson production will be a clear signal of new physics. In the
MSSM with large $\tan\beta$ this can be mainly ascribed to an enhanced
bottom quark Yukawa coupling. In general, however, it cannot be
uniquely interpreted, since other production channels, like the
leading $gg\to h$ gluon fusion , can contribute as well. Detecting one
or two bottom quarks in the final state is enough to remove this
ambiguity, which is the source of the interest in the semi-inclusive
and exclusive production channels, in spite of their smaller cross
section. The exclusive measurement corresponds to the smallest cross
section, but it also has a very reduced background and this is the
case we will consider in the following. The final states can be
further categorized according to the decay of the Higgs boson.
Existing studies have considered mostly the dominant Higgs decay
channel, $h\to b\bar{b}$~\cite{Carena:2000yx,atlas:1999}, but also
$h\to\mu^+\mu^-$~\cite{Dawson:2002cs,Boos:2003jt} and $h\to\tau^+\tau^-$
\cite{Richter-Was:1998ak}.

In view of their phenomenological relevance, a lot of effort has been
recently invested in improving the stability of the theoretical
predictions for the hadronic total cross sections for $p\bar{p},pp\to
t\bar{t}h$ and $p\bar{p},pp\to b\bar{b}h$. Since the tree level or
Leading Order (LO) cross section is affected by a very large
renormalization and factorization scale dependence, the first order or
Next-to-Leading Order (NLO) QCD corrections have been calculated for
the inclusive $p\bar{p},pp\to t\bar{t}h$ cross section
\cite{Beenakker:2001rj,Reina:2001sf,Reina:2001bc,Beenakker:2002nc,Dawson:2002tg,Dawson:2003zu}
and for the the inclusive and exclusive $p\bar{p},pp\to b\bar{b}h$
cross sections
\cite{Dicus:1998hs,Balazs:1998sb,Campbell:2002zm,Dittmaier:2003ej,Dawson:2003pl},
while the inclusive $b\bar{b}\to h$ cross section has also recently
been calculated including the Next-to-Next to Leading Order (NNLO) QCD
corrections \cite{Harlander:2003ai}.  In all cases, the NLO cross
section has a drastically reduced renormalization and factorization
scale dependence, of the order of 15-20\% as opposed to the initial
100\% uncertainty of the LO cross section, and leads to increased
confidence in predictions based on these results. In this proceeding
we will present the results of our calculation of the NLO cross
section for both the inclusive $p\bar{p},pp\to t\bar{t}h$
\cite{Reina:2001sf,Reina:2001bc,Dawson:2002tg,Dawson:2003zu} and the
exclusive $p\bar{p},pp\to b\bar{b}h$ cross sections
\cite{Dawson:2003pl}, where $h$ denotes the SM Higgs boson and, in the
case of $b\bar{b}h$, also the scalar Higgs bosons of the MSSM.

The calculation of the NLO corrections to the hadronic processes
$p\bar{p},pp\to t\bar{t}h$ and $p\bar{p},pp\to b\bar{b}h$ presents
challenging technical difficulties, ranging from virtual pentagon
diagrams with several massive internal and external particles to real
gluon and quark emission in the presence of infrared singularities. A
general overview of the techniques developed and employed in our
calculation are presented in Section~\ref{sec:calculation}, and the
corresponding results are illustrated in Section~\ref{sec:tth} for
$p\bar{p},pp\to t\bar{t}h$ and in Section~\ref{sec:bbh} for
$p\bar{p},pp\to b\bar{b}h$.

\section{THE CALCULATION}
\label{sec:calculation}
The total cross section for $pp\hskip-7pt\hbox{$^{^{(\!-\!)}}$}\to
Q\bar{Q}h$ (for $Q\!=\!t,b$) at ${\cal O}(\alpha_s^3)$ can be written
as:
\begin{eqnarray}\label{eq:sig_nlo}
\lefteqn{\sigma_{\sss NLO}(p\,p\hskip-7pt\hbox{$^{^{(\!-\!)}}$} 
\to Q\bar{Q}h) =\sum_{ij} \frac{1}{1+\delta_{ij} } \int dx_1 dx_2}  \\
&& \hspace*{-0.1cm}\times[{\cal F}_i^p(x_1,\mu) {\cal F}_j^{p(\bar p)}(x_2,\mu)
{\hat \sigma}^{ij}_{\sss NLO}(\mu) + (1\leftrightarrow 2)]\, ,\nonumber
\end{eqnarray}
where ${\cal F}_i^{p(\bar{p})}$ are the NLO parton distribution
functions (PDFs) for parton $i$ in a (anti)proton, defined at a
generic factorization scale $\mu_f\!=\!\mu$, and ${\hat
  \sigma}^{ij}_{\sss NLO}$ is the ${\cal O}(\alpha_s^3)$ parton-level
total cross section for incoming partons $i$ and $j$, made of the
channels $q\bar{q},gg\to Q\bar{Q}h$ and $(q,\bar{q})g\to
Q\bar{Q}h+(q,\bar{q})$, and renormalized at an arbitrary scale $\mu_r$
which we also take to be $\mu_r\!=\!\mu$.  We note that the effect of
varying the renormalization and factorization scales independently has
been investigated and found to be negligible.  The partonic center of
mass energy squared, $s$, is given in terms of the hadronic center of
mass energy squared, $s_{\sss H}$, by $s=x_1 x_2 s_{\sss H}$.

The NLO parton-level total cross section, ${\hat \sigma}^{ij}_{\sss
  NLO}$, consists of the ${\cal O}(\alpha_s^2)$ Born cross section,
${\hat \sigma}^{ij}_{\sss LO}$, and the ${\cal O}(\alpha_s)$
corrections to the Born cross section, $\delta {\hat
  \sigma}^{ij}_{\sss NLO}$, including the effects of mass
factorization. $\delta {\hat\sigma}^{ij}_{\sss NLO}$ contains virtual
and real corrections to the parton-level $Q\bar{Q}h$ production
processes, $q\bar{q}\to Q\bar{Q}h$ and $gg\to Q\bar{Q}h$, and the
tree-level $(q,\bar{q})g$ initiated processes, $(q,\bar{q})g\to
Q\bar{Q}h(q,\bar{q})$, which are of the same order in $\alpha_s$.  

The ${\cal O}(\alpha_s)$ virtual and real corrections to $q\bar{q}\to
Q\bar{Q}h$ and $gg\to Q\bar{Q}h$ have been discussed in detail in
Refs.~\cite{Reina:2001bc,Dawson:2003zu} and we will highlight in the
following only the most challenging tasks.
\subsection{Virtual corrections}
\label{subsec:sigma_virtual}

The calculation of the ${\cal O}(\alpha_s)$ virtual corrections to
$q\bar{q},gg\to Q\bar{Q}h$ (for $Q\!=\!t,b$) proceeds by reducing each
virtual diagram to a linear combination of tensor and scalar
integrals, which may contain both ultraviolet (UV) and infrared (IR)
divergences.  Tensor integrals are further reduced in terms of scalar
integrals~\cite{Passarino:1979jh}.  The finite scalar integrals are
evaluated by using the method described in Ref.~\cite{Denner:1993kt}
and cross checked with the FF package~\cite{vanOldenborgh:1990wn}.
The scalar integrals that exhibit UV and/or IR divergences are
calculated analytically. Both the UV and IR divergences are extracted
by using dimensional regularization in $d\!=\!4-2\epsilon$ dimensions.
The UV divergences are then removed by introducing a suitable set of
counterterms, as described in detail in
Refs.~\cite{Reina:2001bc,Dawson:2003zu,Dawson:2003pl}. The remaining
IR divergences are cancelled by the analogous singularities in the
soft and collinear part of the real gluon emission cross section.

The most difficult integrals arise from the IR-divergent pentagon
diagrams with several massive particles.  The pentagon scalar and
tensor Feynman integrals originating from these diagrams present
either analytical (scalar) or numerical (tensor) challenges. We have
calculated the pentagon scalar integrals as linear combinations of
scalar box integrals using the method of
Ref.~\cite{Bern:1993em,Bern:1994kr}, and cross checked them using the
techniques of Ref.~\cite{Denner:1993kt}.  Pentagon tensor integrals
can give rise to numerical instabilities due to the dependence on
inverse powers of the Gram determinant (GD),
GD$\!=\!\det(p_i\!\cdot\!p_j)$ for $p_i$ and $p_j$ external momenta,
which vanishes at the boundaries of phase space when two momenta
become degenerate. These are spurious divergences, which cause serious
numerical difficulties.  To overcome this problem we have calculated
and cross checked the pentagon tensor integrals in two ways:
numerically, by isolating the numerical instabilities and
extrapolating from the numerically safe to the numerically unsafe
region using various techniques; and analytically, by reducing them to
a numerically stable form.
\subsection{Real corrections}
\label{subsec:sigma_real}
In computing the ${\cal O}(\alpha_s)$ real corrections to
$q\bar{q},gg\to Q\bar{Q}h$ and $(q,\bar{q})g\to Q\bar{Q}h+(q,\bar{q})$
(for $Q\!=\!t,b$) it is crucial to isolate the IR divergent regions of
phase space and extract the corresponding singularities analytically.
We achieve this by using the phase space slicing (PSS) method, in both
the double~\cite{Harris:2001sx} and
single~\cite{Giele:1992vf,Giele:1993dj,Keller:1998tf} cutoff
approaches. In both approaches the IR region of the $Q\bar{Q}h+g$
phase space where the emitted gluon cannot be resolved is defined as
the region where the gluon kinematic invariants:
\begin{equation}
s_{ig}= 2 p_i\cdot p_g=2E_iE_g(1-\beta_i\cos\theta_{ig})
\end{equation}
become small. Here $p_i$ is the momentum of an external (anti)quark or
gluon (with energy $E_i$), $\beta_i\!=\!\sqrt{1-m_i^2/E_i^2}$, $p_g$
is the momentum of the radiated final state gluon (quark/antiquark)
(with energy $E_g$), and $\theta_{ig}$ is the angle between
$\vec{p}_i$ and $\vec{p}_g$.  In the IR region the cross section is
calculated analytically and the resulting IR divergences, both soft
and collinear, are cancelled, after mass factorization, against the
corresponding divergences from the ${\cal O}(\alpha_s)$ virtual
corrections.

The single cutoff PSS technique defines the IR region as that where
\begin{equation}
s_{ig}<s_{min}\,\,\,,
\end{equation}
for an arbitrarily small cutoff $s_{min}$.  The two cut-off PSS method
introduces two arbitrary parameters, $\delta_s$ and $\delta_c$, to
separately define the IR soft and IR collinear regions according to:
\begin{eqnarray}
&&E_g<{\delta_s\sqrt{s}\over 2}\,\,\,\,\,\mbox{soft region}\,\,\,,\nonumber\\
&&(1-\cos\theta_{ig})<\delta_c\,\,\,\,\,\mbox{collinear region}\,\,\,.
\end{eqnarray}

In both methods, the real contribution to the NLO cross section is
computed analytically below the cutoffs and numerically above the
cutoffs, and the final result is independent of these arbitrary
parameters. With this respect, it is crucial to study the behavior of
$\sigma_{\sss NLO}$ in a region where the cutoff(s) are small enough
to justify the analytical calculations of the IR divergent
contributions to the real cross section, but not so small as to cause
numerical instabilities.
\boldmath
\section{RESULTS FOR $t\bar{t}h$ PRODUCTION}
\label{sec:tth}
\unboldmath 
\begin{figure}[tbh]
\begin{center}
\includegraphics[scale=0.42]{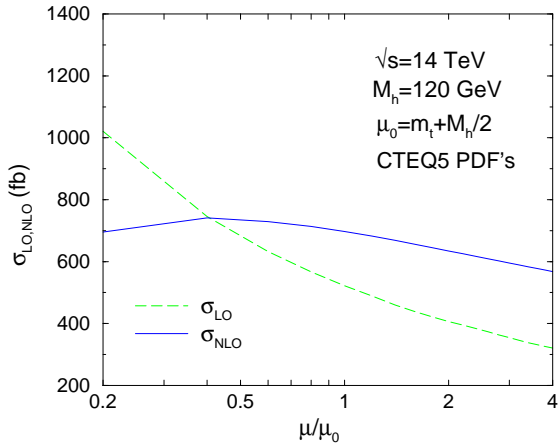}
\vspace{-1.truecm}  
\caption[]{Dependence of $\sigma_{\sss LO,NLO}(pp\to t\bar{t}h)$ on 
  the renormalization/factorization scale $\mu$, at $\sqrt{s_{\sss
      H}}\!=\!14$~TeV, for $M_h\!=\!120$ GeV.}
\label{fg:tth_mudep_lhc}
\end{center}
\vspace{-1.truecm}  
\end{figure}
\begin{figure}[t]
\begin{center}
\includegraphics[scale=0.42]{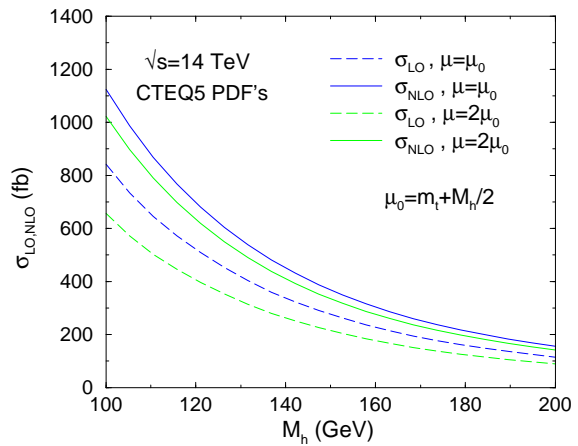}
\vspace{-1.5truecm}  
\caption[]{$\sigma_{\sss NLO}(pp\to t\bar{t}h)$ and $\sigma_{\sss
    LO}(pp\to t\bar{t}h)$ as functions of $M_h$, at $\sqrt{s_{\sss
      H}}\!=\!14$~TeV, for $\mu\!=m_t+M_h/2$ and $\mu\!=\!2m_t+M_h$.}
\label{fg:tth_mhdep_lhc}
\end{center}
\vspace{-1.truecm}  
\end{figure}
The impact of NLO QCD corrections on the tree level cross section for
$pp\to t\bar{t}h$ (LHC) production in the SM is illustrated in
Figs.~\ref{fg:tth_mudep_lhc} and \ref{fg:tth_mhdep_lhc}. Similar
results for the case of $p\bar{p}\to t\bar{t}h$ (Tevatron) can be
found in Ref.~\cite{Reina:2001sf,Reina:2001bc}.  Results for
$\sigma_{\sss LO}$ are obtained using the 1-loop evolution of
$\alpha_s(\mu)$ and CTEQ5L parton distribution functions
\cite{Lai:1999wy}, while results for $\sigma_{\sss NLO}$ are obtained
using the 2-loop evolution of $\alpha_s(\mu)$ and CTEQ5M parton
distribution functions, with $\alpha_s^{\sss NLO}(M_Z)\!=\!0.118$. The
top quark mass is renormalized in the OS scheme and its pole mass is
fixed at $m_t\!=\!174$~GeV.

Fig.~\ref{fg:tth_mudep_lhc} illustrates the
renormalization/factorization scale dependence of $\sigma_{\sss LO}$
and $\sigma_{\sss NLO}$ at the LHC. The NLO cross section shows a
drastic reduction of the scale dependence with respect to the lowest
order prediction.  Fig.~\ref{fg:tth_mhdep_lhc} complements this
information by illustrating the dependence of the LO and NLO cross
sections on the Higgs boson mass at the LHC. For scales
$\mu\!\ge\!0.4\mu_0$ ($\mu_0\!=\!m_t+M_h/2$) the NLO corrections
enhance the cross section. We estimate the remaining theoretical
uncertainty on the NLO result to be of the order of 15-20\%, due to
the left over $\mu$-dependence, the error from the PDFs, and the the
error on the top quark pole mass $m_t$.

\boldmath
\section{RESULTS FOR $b\bar{b}h$ PRODUCTION}
\label{sec:bbh}
\unboldmath We evaluate the fully exclusive cross section for
$b\bar{b}h$ production by requiring that the transverse momentum of
both final state bottom and anti-bottom quarks be larger than 20~GeV
($p_T^b\!>\!20$~GeV), and that their pseudorapidity satisfy the
condition $|\eta_b|\!<\!2$ for the Tevatron and $|\eta_b|\!<\!2.5$ for
the LHC.  This corresponds to an experiment measuring the Higgs decay
products along with two high $p_T$ bottom quark jets. In order to
better simulate the detector response, the final state gluon and the
bottom/anti-bottom quarks are treated as distinct particles only if
the separation in the azimuthal angle-pseudorapidity plane is $\Delta
R\!>\!0.4$.  For smaller values of $\Delta R$, the four momentum
vectors of the two particles are combined into an effective
bottom/anti-bottom quark momentum four-vector.

As for $t\bar{t}h$ production, our numerical results for the NLO (LO)
cross sections are obtained using CTEQ5M (CTEQ5L) PDFs
~\cite{Lai:1999wy} and the 2-loop (1-loop) evolution of
$\alpha_s(\mu)$ . In the $b\bar{b}h$ case, however, we have also
investigated the dependence of the NLO result on the choice of the
renormalization scheme for the bottom quark Yukawa coupling.  The
strong scale dependence of the $\overline{MS}$ bottom quark mass
($\overline{m}_b(\mu)$) plays a special role in the perturbative
evaluation of the $b\bar{b}h$ production cross section since it enters
in the overall bottom quark Yukawa coupling.  The same is not true for
$t\bar{t}h$ production since the $\overline{MS}$ top quark mass has
only a very mild scale dependence.  The bottom quark pole mass is
taken to be $m_b\!=\!4.6$~GeV. In the OS scheme the bottom quark
Yukawa coupling is calculated as $g_{b\bar{b}h}\!=\!m_b/v$, while in
the $\overline{MS}$ scheme as
$g_{b\bar{b}h}(\mu)\!=\!\overline{m}_b(\mu)/v$, where we use the
2-loop (1-loop) $\overline{MS}$ bottom quark mass for the NLO (LO)
cross section respectively.
\begin{figure}[bth]
\begin{center}
\includegraphics[scale=0.42]{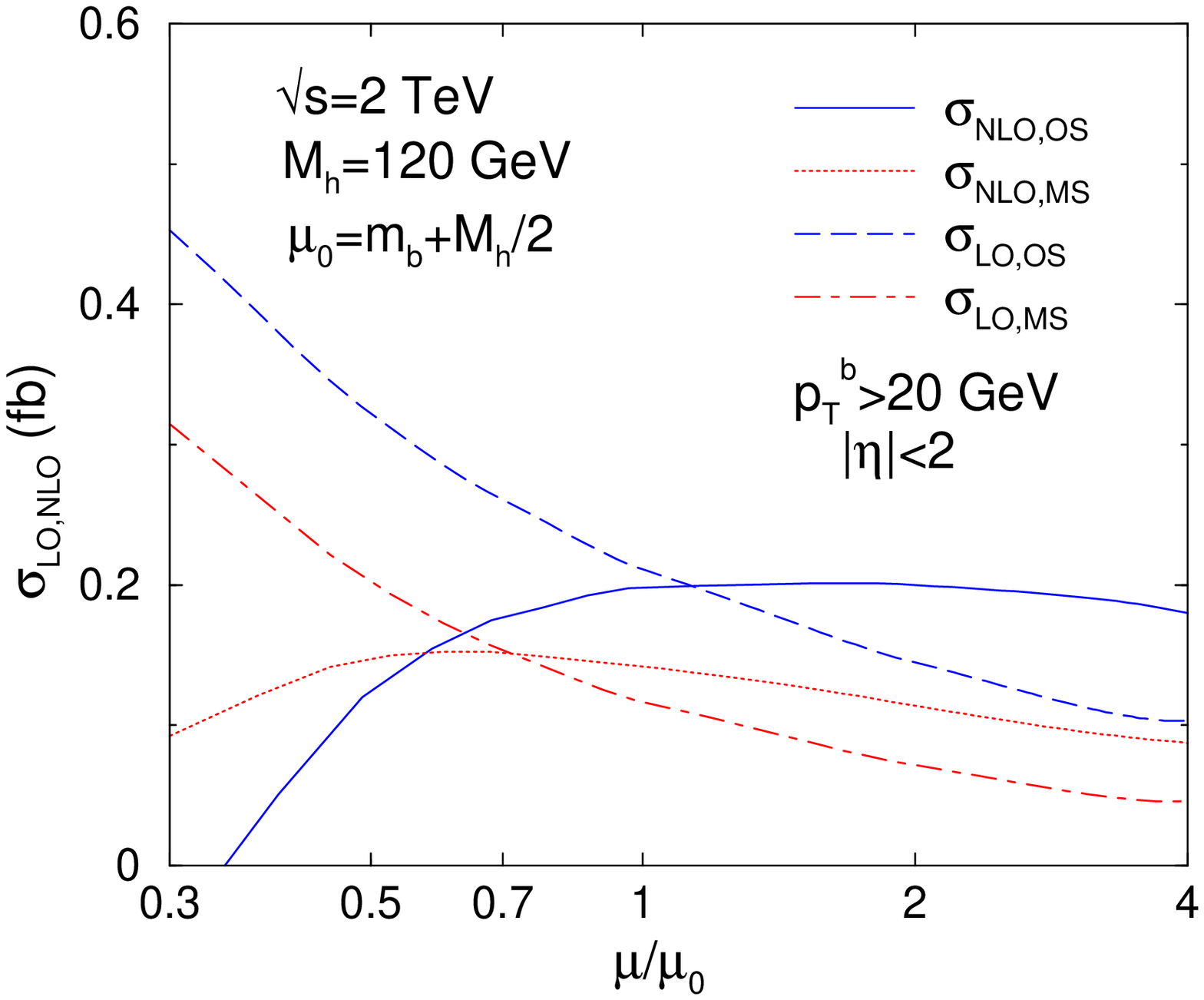} 
\includegraphics[scale=0.42]{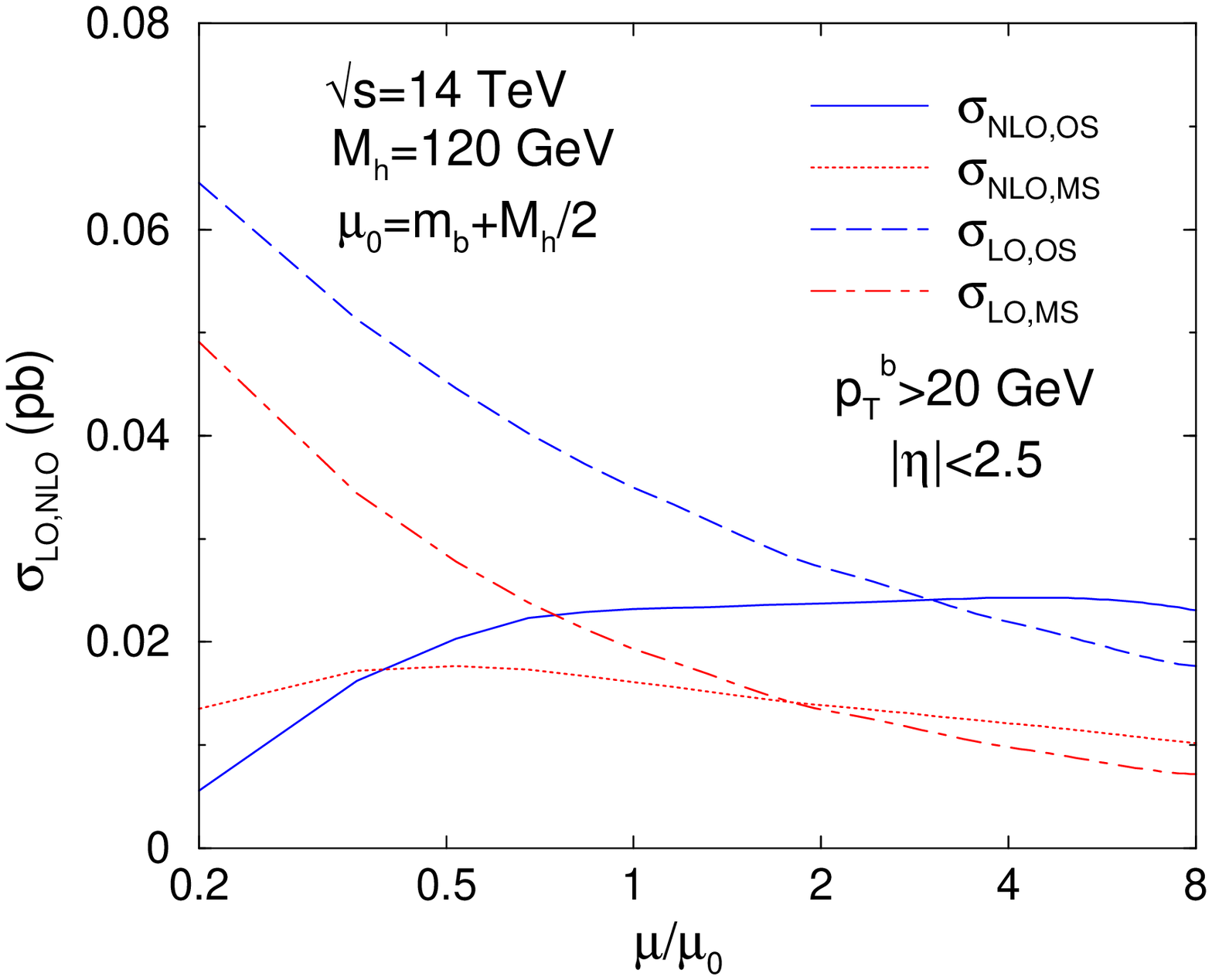} 
\vspace{-1.2truecm}
\caption[]{$\sigma_{\sss NLO}$ and $\sigma_{\sss LO}$ for $p\bar{p}\to
  b\bar{b}h$ at $\sqrt{s}\!=\!2$~TeV (top) and for $pp\to b\bar{b}h$
  at $\sqrt{s}\!=\!14$~TeV (bottom) as a function of the
  renormalization/factorization scale $\mu$, for $M_h=120$~GeV.}
\label{fg:bbh_mu_dep}
\end{center}
\vspace{-1.truecm}
\end{figure}

The impact of NLO QCD corrections on the tree level cross section for
$b\bar{b}h$ exclusive production in the SM is summarized in
Fig.~\ref{fg:bbh_mu_dep} for both the Tevatron and the LHC. In both
the $OS$ and the $\overline{MS}$ schemes the stability of the cross
section is greatly improved at NLO, and the correspondent theoretical
uncertainty reduced to 15-20\%. The $\overline{MS}$ results seem to
have overall a better perturbative behavior, although the variation of
the NLO cross section about its point of least sensitivity to the
renormalization/factorization scale is almost the same when one uses
the $OS$ or $\overline{MS}$ schemes for the bottom Yukawa coupling.
This indicates that the running of the Yukawa coupling is not the only
important factor to determine the overall perturbative stability of
the cross section. The difference between the $OS$ and $\overline{MS}$
results at their plateau values should probably be interpreted as an
additional theoretical uncertainty.

\begin{figure}[t]
\begin{center}
\includegraphics[scale=0.42]{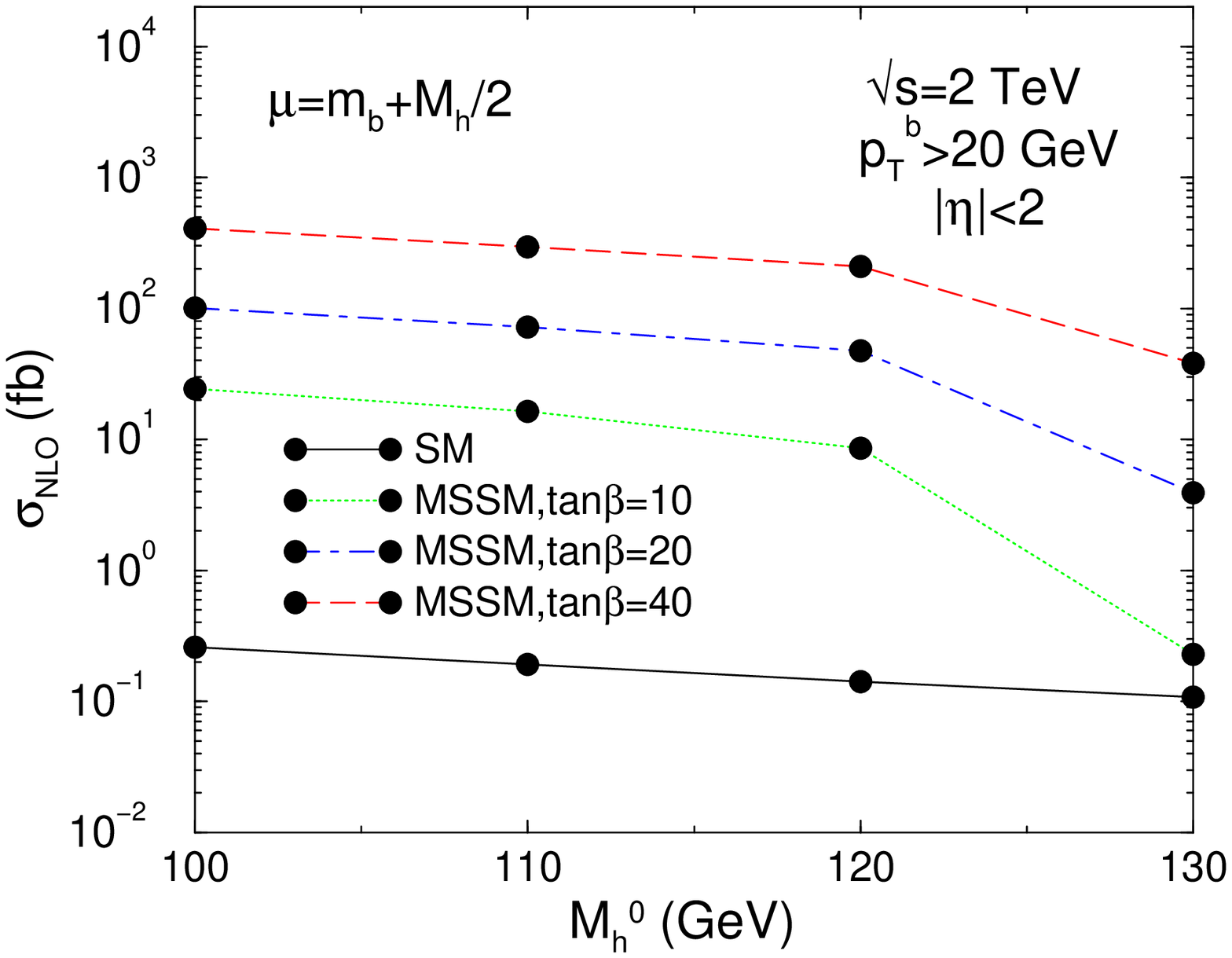} 
\includegraphics[scale=0.42]{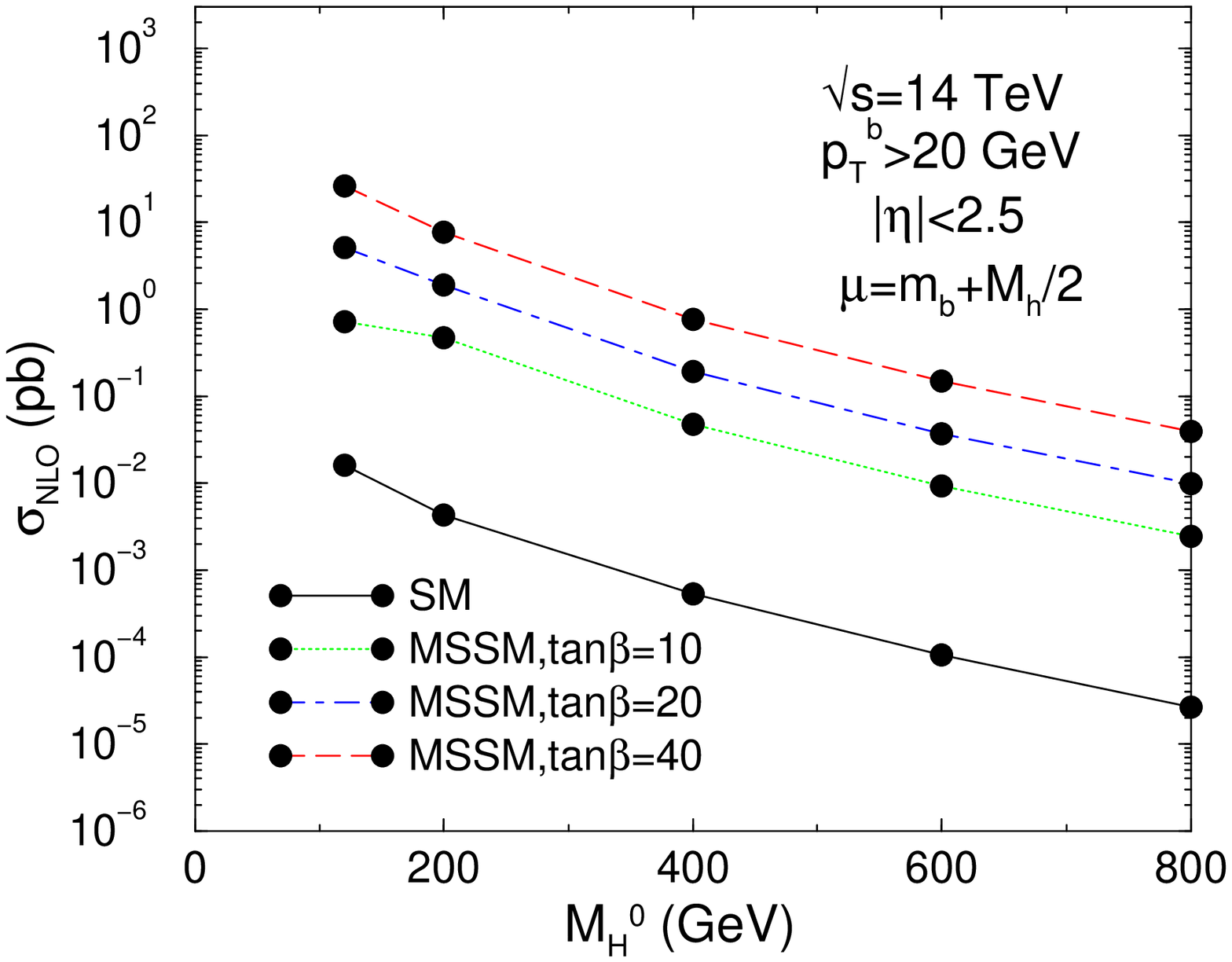} 
\vspace{-1.2truecm}
\caption[]{ $\sigma_{\sss NLO,MS}$ 
  for $p\bar{p}\to b\bar{b}h$ production at $\sqrt{s}\!=\!2$~TeV (top)
  and $pp\to b\bar{b}h$ production at $\sqrt{s}\!=\!14$~TeV (bottom)
  in the SM and in the MSSM with $\tan\beta\!=\!10,20$, and $40$.}
\label{fg:bbh_mh_dep}
\end{center}
\vspace{-1.truecm}
\end{figure}

Finally, in Fig.~\ref{fg:bbh_mh_dep} we illustrate the dependence of
the exclusive cross section, at the Tevatron and at the LHC, on the
Higgs boson mass, both in the SM and in some scenarios of the MSSM,
corresponding to $\tan\beta\!=\!10,20$ and $40$. For the Tevatron we
consider the case of the light MSSM scalar Higgs boson($h^0$) while
for the LHC we consider the case of the heavy MSSM scalar Higgs boson
($H^0$). We see that the rate for $b\bar{b}h$ production can be
significantly enhanced in a supersymmetric model with large values of
$\tan\beta$, and makes $b\bar{b}h$ a very important mode for discovery
of new physics at both the Tevatron and the LHC.
\section*{Acknowledgments}
The work of S.D. (C.J., L.H.O., L.R.)  is supported in part by the
U.S.  Department of Energy under grant DE-AC02-98CH10886
(DE-FG02-97ER41022, DE-FG-02-91ER40685, DE-FG02-97ER41022). The work
of D.W. is supported in part by the National Science Foundation under
grant No.~PHY-0244875.
\bibliography{qcd03}
\end{document}